\newcommand{\be}{\begin{equation}}
\newcommand{\ee}{\end{equation}}

\documentclass[aps,prl,twocolumn]{revtex4}
\usepackage{graphicx}
\usepackage{amssymb}
\usepackage{bm}

\begin{document}

\title{Matter-wave gap vortices in optical lattices}

\author{Elena A. Ostrovskaya and Yuri S. Kivshar}

\affiliation{Nonlinear Physics Group and ARC Centre of Excellence for 
Quantum-Atom Optics, Research School of Physical Sciences and
Engineering, Australian National University, Canberra ACT 0200,
Australia}


\begin{abstract}
We predict the existence of spatially localized nontrivial topological states of a Bose-Einstein condensate with repulsive atomic interactions confined by an optical
lattice. These nonlinear localized states, {\em matter-wave gap vortices}, possess a vortex-like phase dislocation and exist in the {\em gaps} of the matter-wave bandgap spectrum due to the Bragg scattering. We discuss the structure, stability, and formation dynamics of the gap vortices in the case of two-dimensional optical lattices.
\end{abstract}
\maketitle

Similarities between the physics of coherent light and matter
waves can be successfully used to understand and predict nonlinear
dynamics of weakly interacting Bose-Einstein condensates (BECs) in
optical lattices~\cite{pu,our_prl,marcus_dm,inguscio_dm,optex}. In
analogy with periodic photonic structures for light waves, such as
photonic crystals, optical lattices form band-gap structures which
modify diffraction and localization properties of BECs with both attractive and repulsive
atomic interactions. In the case of a repulsive BEC,
optical lattices offer {\em two intriguing possibilities}. First,
they enable nonlinear localization of a repulsive condensate in the gaps of the Bloch-wave
spectrum, without a confining harmonic potential  ~\cite{Zob_et99,lenaPRA}. 
Secondly, by manipulating the band gaps of the lattice Bloch-wave spectrum, the
degree of BEC localization can be varied from a low density
state spread out across the lattice to a condensate droplet
tightly bound in a vicinity of a single lattice well. Both the BEC
diffraction management in shallow one-dimensional (1D) optical
lattices~\cite{marcus_dm,inguscio_dm} and the nonlinear localization of a BEC with repulsive
 atomic interactions~\cite{gap_observation} were observed in experiments.

One of the fundamental questions is how distinctively superfluid
properties of weakly interacting condensates, e.g., the dynamics of
 vortices~\cite{bec_vortices}, are modified by an optical lattice, and whether it is possible to
spatially localize a BEC wavepacket with a nontrivial topological
phase in a lattice. Several parallel studies in the physics of
coherent optical~\cite{yang} and matter waves~\cite{malomed_el}
suggest that  localization of topological defects is possible for an {\em attractive}
condensate within a {\em tightly binding} lattice. The localized
vortices in the attractive condensate would be analogous to bright
optical vortex solitons predicted to exist in total internal
reflection gaps of dynamical photonic lattices~\cite{yang,ta} and
photonic crystal fibers \cite{ph} with {\em focusing} nonlinearity, and recently observed in
experiments on photonic lattices~\cite{neshev}.
Localization of vortices in the {\em repulsive} BEC confined by a {\em shallow}
lattice, i.e.  when the standard mechanisms of spatial confinement are {\em absent}, remains an open problem.

In this Letter we show that
nonlinear localization of a repulsive BEC with a phase
defect is possible within a complete Bragg-reflection gap of the
Bloch-wave spectrum. The resulting {\em matter-wave gap vortices} are
spatially localized and dynamically stable. Due to their
finite spatial extent, which is much smaller than the
dimensions of the loose confining trap, the localized gap vortices
can also be termed ``bright" vortices, by analogy with vortex
rings of light in nonlinear optical media. Surprisingly, the
density and phase structure of a localized lattice vortex
is determined by the features of delocalized "dark" vortices,
which reside on a spatially extended periodic background of
matter-wave Bloch states.


The framework of our analysis is set by the mean-field
Gross-Pitaevskii model of an anisotropic BEC cloud loaded into a
two-dimensional optical lattice potential,
\begin{equation}
\label{dimless} \label{eq2D} i\frac{\partial\Psi}{\partial t} =
\left\{-\nabla_{\perp}^{2} + V(x,y) +
 |\Psi|^{2}\right\}\Psi.
\end{equation}
This equation is obtained by assuming a tight confinement in the
direction perpendicular to the lattice ("pancake" trapping geometry)
and a standard dimensionality reduction procedure (see, e.g., Ref.
\cite{our_prl}). It is made dimensionless by using the
characteristic length $a_{L}=d/\pi$, energy
$E_{L}=\hbar^{2}/2ma_{L}^{2}$, and time
$\omega^{-1}_{L}=\hbar/E_{L}$ scales of the lattice, where $d$
is the lattice period, and $m$ the mass of the trapped atoms. The
wavefunction is scaled as $\Psi \to \Psi \sqrt{g_{\rm 2D}}$, where
the rescaled 2D interaction strength is $g_{\rm
2D}=4\sqrt{2}\pi a_s/a_L$, and $a_s$ is the s-wave scattering length. 
To eliminate any possible localization effects which are {\em not} 
due to the optical lattice, we {\em
removed} from  consideration any additional trapping of the
condensate in the directions of the lattice. The trapping
potential is therefore taken in the form, $ V(x,y) = V_0 [
\sin^{2}(x+\theta)+ \sin^{2}(y+\theta)]$.

\begin{figure}
\centerline{\scalebox{0.7}{\includegraphics{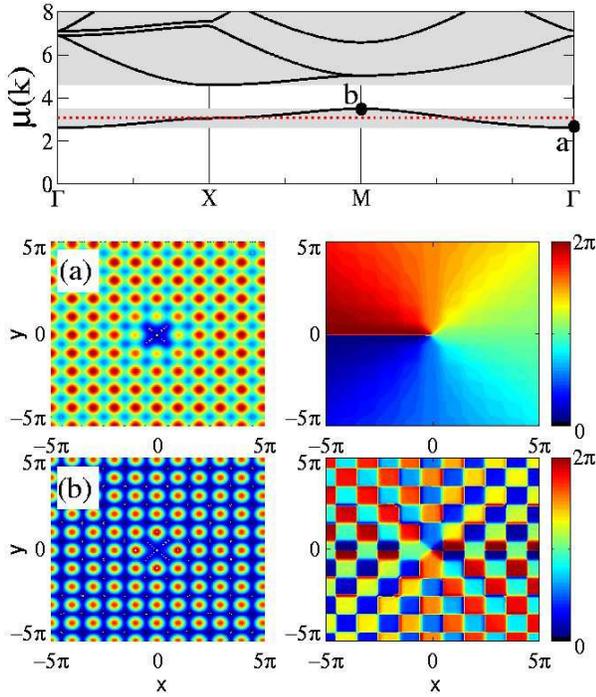}}}
\caption{ Top: Matter-wave Bloch spectrum in a 2D
optical lattice ($V_0=3.0$) shown along the edges of the
irreducible BZ, $\Gamma(k_x=0,k_y=0) \to
X(k_x=1,k_y=0) \to M(k_x=1,k_y=1)$. Shaded and open areas show bands and gaps, respectively. Bottom: examples of the condensate wavefunctions, $|\psi|^2$, (red color codes maximum density)  at the
marked points of the dispersion curves, corresponding to (a)
off-site vortex in the middle of the first BZ, and (b)
on-site vortex at the edge of the first BZ, shown
together with their phase structures. } \label{fig1}
\end{figure}

Stationary states of a BEC are described by solutions
of Eq.~(\ref{eq2D}) of the form: $\Psi({\bf r},t) = \psi({\bf
r})\exp(-i\mu t)$, where $\mu$ is the chemical
potential. Stationary states of the non-interacting BEC are found
as $\psi({\bf r})=\phi_{{\bf k}}({\bf r})\exp(i {\bf k}{\bf r})$,
where the wavevector ${\bf k}$ belongs to the Brillouine  zone (BZ) of
the square lattice, and $\phi_{\bf k}({\bf r})=\phi_{\bf k}({\bf
r}+{\bf d})$ is a periodic (Bloch) function with the periodicity
of the lattice. The band-gap structure of spectrum, $\mu(\bf{k})$, of the atomic Bloch
waves in the optical lattice is shown in the top panel of Fig.~\ref{fig1}, in the
reduced zone representation usually employed in the theory of
crystalline solids.

In a homogeneous BEC, a vortex with the topological charge one is characterized by a $2\pi$ phase winding around the low density core on a constant density background. In a condensate loaded into an "infinite"  optical lattice,  a single-charge vortex can be stabilized in the form of a "dark" {\em Bloch vortex} on a periodic background of a Bloch state. Two examples of stationary
Bloch vortices found by numerical solution  \cite{our_prl, jj} of the time-independent
version of Eq. (1) are shown in Fig.~\ref{fig1}(a,b).  Bloch
vortices on the background of the ground lattice state [i.e. at
the lowest edge of the first band, Fig.~\ref{fig1}(a)] have the
winding structure of the phase typical for a charge one vortex in a homogeneous condensate,
due to the {\em trivial phase} of the background Bloch state
\cite{pu,optex}. In contrast, the Bloch vortex residing on the
excited Bloch state at the edge of the 1-st BZ with {\em
nontrivial phase} develops a highly unusual phase pattern [Fig. 1(b)].
The phase winding around the zero density centre is 
$2\pi$.

The ``dark'' vortices on the ground (Bloch) state were
found to be dynamically stable in~\cite{arimondo,malomed}, although their experimental
observation could be difficult because their core size is of the order of the lattice period. Recent analysis of the Bose-Hubbard model for a strongly correlated bosonic system with nearest neighbor repulsion in an optical
lattice~\cite{wu} also revealed the existence of a single vortex with
the particle density that can be either suppressed or {\em
enhanced} in the vortex core.

\begin{figure}
\centerline{\scalebox{0.38}{\includegraphics{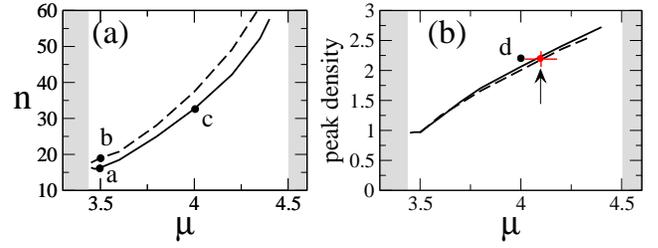}}}
\caption{Variations of (a) normalized atom number and (b) 
peak density of the on-site (solid) and off-site (dashed) gap
vortex within the complete gap (open area) at $V_0=3$. Shaded areas are spectral bands. Examples of the spatial structure of the vortices at the marked points 
are shown in Fig. 3.} \label{fig2}
\end{figure}

\begin{figure}
\centerline{\scalebox{0.7}{\includegraphics{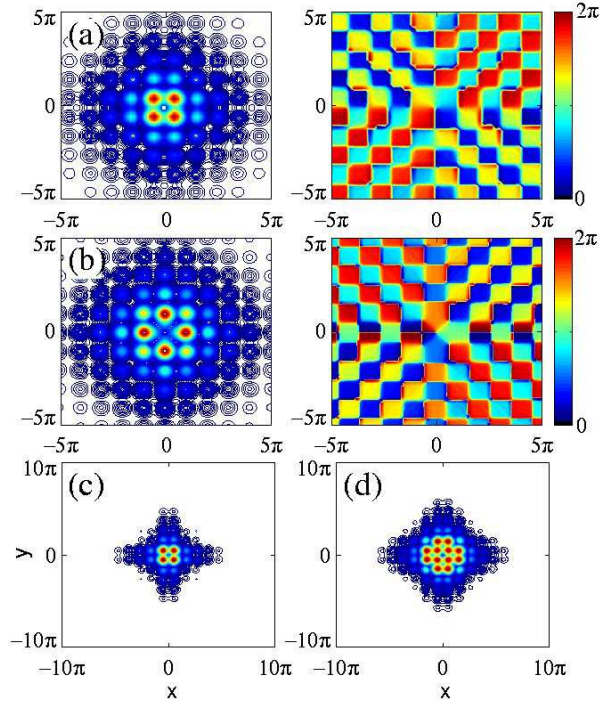}}}
\caption{ (a,b) Density, $|\psi|^2$, and phase
structure of the off-site (a) and on-site (b) gap vortices at marked points on
the existence curves in Fig.~\ref{fig2}.
(c) Typical profile of a strongly localized off-site vortex (at
$\mu=4.0$). (d) Example of a strongly localized broad off-site
vortex (at $\mu=4.0$, $n=124$), with the parameters off the
scale of Fig.~\ref{fig2}(a), but peak density shown in
Fig.~\ref{fig2}(b), point ``d''. Red color codes maximum density.} \label{fig3}
\end{figure}

Quite apart from the spatially delocalized states within the Bloch
bands described above, {\em spatially localized} states of BEC with
repulsive interactions can only exist within the complete spectral gaps~\cite{our_prl,optex}. 
By imposing the vortex-like phase structure onto a spatially
localized envelope of a  Bloch state at the $M$-edge of the
spectral gap (i.e. in the vicinity of the point $b$ in Fig. 1),  we have found, numerically, different families of spatially localized "bright" matter-wave gap vortex solitons characterized
by the dependence of the (normalized) number of atoms in the
localized state, $n=\int |\psi|^2 dx dy$, on the chemical potential [see
Fig.~\ref{fig2}(a)].  Akin to the bright gap matter-wave solitons \cite{salerno}, gap vortices have a clear atom number cut-off below which they undergo a delocalizing transition. The lowest energy families shown in Fig.~\ref{fig2} correspond to the on-site (localized on the
lattice minimum) and off-site (localized on the lattice maximum)
stationary gap vortices shown in Figs.~\ref{fig3}(a,b) respectively.
The peak density of the localized state is a function of chemical
potential within the gap, as shown in Fig.~\ref{fig2}(b), however
the width (FWHM) of the high-density vortex core practically does not
change across the gap. Gap vortices are strongly localized in the
middle of the gap [see Fig.~\ref{fig3}(c)], but near the top ($X$) edge
of the complete gap they develop extended tails that have the amplitude and
phase structure of the background Bloch state.

Strongly localized core of the gap vortex  is similar to vortex "cells" that were predicted to exist in tightly binding photonic lattices optically-induced in a {\em focusing} nonlinear medium
\cite{yang}. Here, we find that the "elemetary" vortex cells of two
basic types form broader gap vortices by "tiling" the square lattice in
a symmetric manner [c.f. Figs. 3 (c) and (d)]. The width of the gap vortex core is 
an additional parameter which characterizes higher-order
families of the vortex states, however the peak density, $n_{\rm max}\equiv|\psi|^2_{\rm max}$,
of the vortex core varies very little
between "elementary" and broad vortices [see point
$d$ in Fig.~\ref{fig2}(b) indicating the peak density of the vortex in Fig. 3(d)]. The high-density core of a broad gap vortex preserves characteristic phase structure of a vortex "cell", with
its $2\pi$ winding structure, whereas its tails are characterized
by a {\em nontrivial phase pattern} associated with the nontrivial
phase of the underlying Bloch state, as seen in
Figs.~\ref{fig3}(a,b).

The localized states near the $M$ edge of the complete gap can be 
described by the trial function:
\begin{equation}
\label{ansatz} \psi(x,y)=Ar^m \, \exp (-r^2/a^2+i\phi) \, B_1(x,y),
\end{equation}
where $r^2=x^2+y^2$,  and $B_1(x,y)$ is the Bloch state at the
lower edge of the first gap.  The Bloch state at the respective
edge is always out-of-phase with the lattice potential, and due to
separability of the lattice potential it can be well approximated as
$B_1\approx\cos(x+\theta)\, \cos(y+\theta)$ (see, e.g. \cite{pu}), where
$\theta$ is a shift of the vortex center relative to the lattice
 minimum.  The approximate form of the on-site and
off-site vortex configurations is well captured by the trial
function (\ref{ansatz}) with $m=2$, $\theta=0$, and  $m=1$, $\theta=\pi/2$,
respectively. In both cases  the necessary condition for the
existence for a stationary state, i.e. zero linear momentum
\cite{anton} $\int Im({ \psi^*} \nabla \psi) dx dy=0$, is
satisfied in the limit of $a\gg 1$.  The characteristic width of
the gap vortex is given by $a=p\pi$, where integer $p$ determines
the size of the high-density vortex core in lattice periods. The
examples of exact numerical solutions of well-localized off-site
vortices corresponding to $p=1$ and $p=2$ are presented in
Figs.~\ref{fig3}(c,d), respectively.

\begin{figure}
\centerline{\scalebox{0.5}{\includegraphics{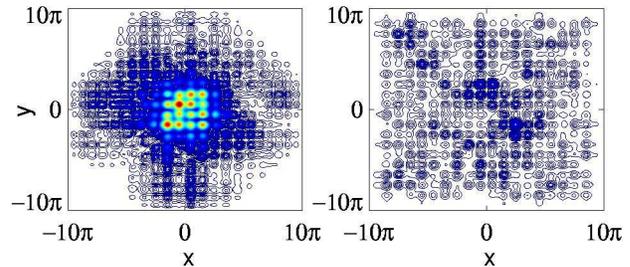}}}
\caption{Snapshots of the density distribution for
a dynamically unstable  wide off-site vortex in Fig. 3(d) at 
evolution times t=252 (left) and t=600 (right).} \label{fig4}
\end{figure}

By direct numerical simulations of Eq. (1) we have confirmed that both
the off-site and on-site gap vortices
can be dynamically stable. The phase distribution evolves
and the phase dislocation {\em precesses} around the centre of the
vortex, whereas the vortex magnitude and polarity of its charge
are {\em preserved} in time. In contrast, broad gap vortices characterized by much higher atom
numbers in the localized state are {\em dynamically unstable}, and
 experience rapid delocalization at relatively long evolution
times, as shown in Fig.~~\ref{fig4}. This process also destroys
the phase structure of the gap vortex core.

Finally, we discuss the feasibility of dynamical generation of gap vortices in an
experiment. To date, generation of nonlinear localized states
of BEC in the form of a {\em bright gap soliton} has been successfully
achieved in a 1D optical lattice \cite{gap_observation}, near
the edge of the first BZ. The key to the success of that
experiment was a low atom number ($N\sim 10^2$) that allowed
generation of a fundamental gap soliton near the band edge. In
contrast to the 1D case, the existence of the particle number
threshold for the localized states in 2D \cite{salerno} means that the
generation process can fail for insufficient initial atom numbers \cite{dudarev}.  
Our analysis shows that
the key ingredients to successful generation of the gap vortex
are: (i) preparation of the BEC wavepacket at the edge $M$ of the
first BZ, (ii) initial number of atoms well exceeding the threshold
atom number for gap vortices, and (iii) the peak density of the
BEC wavepacket being above the threshold peak density for a gap
vortex. While the requirements (i) and (ii) are
quite intuitive, the requirement (iii) is most stringent. To
demonstrate the dramatic role of the peak density thresholds, we impose a $2\pi$ phase winding ramp onto
a broad wavepacket near the edge of the first BZ and explore its evolution
in the dynamical simulations of the Eq. (1). We therefore assume
that, initially, the wave function has a form given by Eq.
(\ref{ansatz}) with $m,\theta=0$, and a typical density
distribution shown in Fig. 5(a). The numerical simulations of the
mean-field model in \cite{dudarev} suggest that the preparation
(and localization) of the initial BEC wavepacket in the above
form can be achieved via an adiabatic
process of driving condensate to the $M$ edge of the BZ with
subsequent expansion in the lattice. The phase dislocation can be
subsequently imprinted by a laser field \cite{phase_imprinting}.
The results of the numerical simulations for a shallow lattice ($V_0=3.0$), with $a=4.5\pi$ and
$A=1.2$, $0.7$, and $0.6$ are presented in top panel of Fig. 5 by curves (b), (c), and
(d), respectively.  In the cases
(b), (c), and (d) the initial number of atoms exceeds the
threshold atom number value for generation of an on-site gap
vortex by $9$, $2$, and $1.7$ times, respectively.  However, the
initial peak density in the case (b) lies {\em above} the gap vortex
threshold, in the case (c) it is in the {\em intermediate} domain
(shaded area in Fig. 5, top) between the thresholds for a gap vortex
and a bright gap soliton, and in the case (d) the peak density is
only {\em marginally below} the threshold for generation of a
bright gap soliton.

\begin{figure}
\centerline{\scalebox{0.45}{\includegraphics{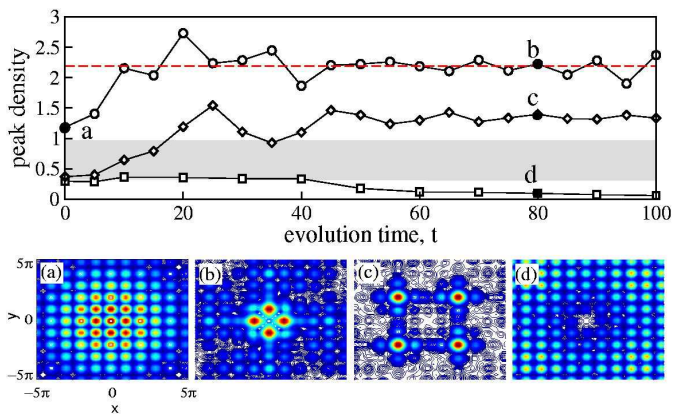}}}
\caption{Top: Evolution of the peak density of a
BEC wavepacket
with the initial peak density  (circles) above threshold for generation of the gap
vortices, (diamonds) below the gap vortex threshold but above the
bright gap soliton threshold,  and  (squares) below bright gap
soliton threshold ($V_0=3.0$). Bottom: Snapshots of the condensate
density at marked points: (a) an initial wavepacket, (b) generated on-site gap vortex with the phase
structure (not shown) of Fig. 3(b), (c) generated weakly
bound steady state of four bright gap solitons, and (d) discrete
diffraction.} \label{fig5}
\end{figure}

The results of the initial state evolution are shown in Fig. 5,
bottom. In the case (b), the on-site gap vortex is generated in
the transient process where the BEC density distribution is
evolving around a vortex state with the peak density of
$n_{\rm max}\approx2.2$, marked by a dashed line in Fig. 5. The one-to-one
correspondence between the peak density of a stationary gap vortex
state and its chemical potential [see Fig. 2(b)] allows us to
determine the approximate chemical potential (with standard
deviation) of the generated
on-site vortex state [pointed at by an arrow in Fig. 2(b)], and identify it as a {\em well-localized gap state}.
Only approximately $40\%$ of the initial atom numbers are retained
in the localized state. Growth in the initial number of
atoms, i.e. due to a greater width of the initial
wavepacket, leads to the increasing deviations from the average
state, and to larger uncertainties in both $\mu$
and $n_{\rm max}$ of the generated vortex. In the case shown in Fig. 5(c)  a
weakly bound state of bright gap vortices forms, with an arbitrary
phase relationship between the soliton peaks. The retention of
atoms in the localized states is almost $90\%$. In the case (d)
the localized states do not form and instead the discrete diffraction of
the initial state is observed.


In conclusion, we have predicted novel types of localized vortex
states of repulsive BECs in optical lattices, and
identified the key requirements for their generation. Our results
provide the next step in understanding the complex interplay
between the superfluid behavior of a BEC with its inherent
signatures, such as vorticity, and the nonlinear behavior of
coherent matter waves in periodic potentials. The experimental observation of predicted structures seems feasible in view of the recent observation of  a 2D BEC in an optical surface trap, and the fascinating prospects of the creation of optical surface lattices \cite{grimm}. Our predictions are also  relevant to the light beams carrying phase dislocations in 2D photonic structures with a defocusing nonlinearity.


\begin{thebibliography}{}

\bibitem{pu} H. Pu {\em et al.}, Phys. Rev. A  {\bf 67}, 043605 (2003).

\bibitem{our_prl} E.A. Ostrovskaya and Yu.S. Kivshar, Phys. Rev.
Lett. {\bf 90}, 160407 (2003).

\bibitem{marcus_dm} B. Eiermann {\em et al.}, Phys. Rev. Lett. {\bf 91}, 060402
(2003).

\bibitem{inguscio_dm} L. Fallani {\em et al.}, Phys. Rev. Lett. {\bf 91}, 240405 (2003).

\bibitem{optex} E.A. Ostrovskaya and Yu.S. Kivshar, Opt. Express {\bf 12}, 19 (2004).

\bibitem{Zob_et99} O. Zobay {\em et al.}, Phys. Rev. A {\bf 59}, 643 (1999).

\bibitem{lenaPRA} P.J. Louis {\em et al.}, Phys. Rev. A {\bf 67}, 013602 (2003).

\bibitem{gap_observation} B. Eiermann {\em et al.}, arXiv:cond-mat/0402178 (2004).

\bibitem{bec_vortices} J.E. Williams and M.J. Holland, Nature (London)  {\bf 401}, 586 (1999);
M.R. Matthews {\em et al.}, Phys. Rev. Lett. {\bf 83}, 2498 (1999).

\bibitem{yang} J. Yang and Z. H. Musslimani, Opt. Lett. {\bf 28}, 2094 (2003); J. Yang, arXiv:nlin.PS/0310024 (2003).

\bibitem{malomed_el} B.B. Baizakov {\em et. al.}, Europhys. Lett. {\bf 63}, 642 (2003).

\bibitem{ta} T.J. Alexander {\em et al.}, arXiv:physics/0403103 (2004).

\bibitem{ph} A. Ferrando {\em et al.}, Opt. Express {\bf 12}, 817 (2004).

\bibitem{neshev} D. N. Neshev {\em et al.}, Phys. Rev. Lett. 92, 123903 (2004).

\bibitem{jj} J.J. Garc\`{\i}a-Ripoll and V.M. P\`erez-Garc\`{\i}a, SIAM J. Sci. Comput. {\bf 23}, 1316 (2001).

\bibitem{arimondo} A. B Bhattacherjee {\em et al.}, arXiv:cond-mat/0312124 (2003).

\bibitem{malomed} P.G. Kevrekidis {\em et al.}, J. Phys. B, {\bf 36}, 3467 (2003).

\bibitem{wu} C. Wu {\em et al.},  Phys. Rev. A {\bf 69}, 043609 (2004).

\bibitem{salerno} see B. B. Baizakov and M. Salerno, Phys. Rev. A, {\bf 69}, 013602 (2004); N. K. Efremidis {\em et al.}, Phys. Rev. Lett., {\bf 91}, 213906 (2003), and references therein.

\bibitem{anton} A. S. Desyatnikov and Yu. S. Kivshar, J. Opt. B, {\bf 4}, 58 (2002).

\bibitem{dudarev} A. M. Dudarev {\em et al.}, arXiv:cond-mat/0312054 (2003).

\bibitem{phase_imprinting} L. Dobrek et al., Phys. Rev. A {\bf 60}, R3381 (1999).

\bibitem{grimm} D. Rychtarik {\em et al.}, Phys. Rev. Lett. {\bf 92}, 173003 (2004).


\end{thebibliography}
\end{document}